\documentclass[12pt,a4paper]{article}

\usepackage{longtable} 

\usepackage{hyperref}
\usepackage{color}
\usepackage{float}
\floatstyle{plaintop}
\restylefloat{table}
\usepackage{fourier} 
\usepackage{array}
\usepackage{makecell,adjustbox}

\usepackage{graphicx}
\usepackage{sidecap} 
\usepackage{sidecap}
\usepackage{caption}
\usepackage[font=small,labelfont=bf]{caption}

\renewenvironment{abstract}
 {\small
  \begin{center}
  \bfseries \abstractname\vspace{-.5em}\vspace{0pt}
  \end{center}
  \list{}{%
    \setlength{\leftmargin}{5mm}
    \setlength{\rightmargin}{\leftmargin}%
  }%
  \item\relax}
 \endlist 

\title{
Proton radius reconstruction from simulated electron-proton elastic scattering cross sections at low transfer momenta}

\author{
S.Belostotski,  N.Sagidova,  A.Vorobyev
 \bigskip\\
\normalfont\itshape
Petersburg Nuclear Physics Institute\\ 
\normalfont\itshape\small
NRC "Kurchatov Institute", Gatchina, Russia\\
\normalfont\footnotesize  Contact person : Alexey Vorobyev \\
\normalfont\footnotesize email : {vorobyov\_aa@pnpi.nrcki.ru}
}

\begin{document}

\maketitle

\begin{abstract}

     This note is motivated by preparations of a  new $ep$  elastic scattering experiment  in the low transfer momentum region to be carried out  in the 720 MeV electron beam of the Mainz Microtron MAMI. This experiment will use an innovative method allowing for detection of recoil protons in coincidence  with the scattered electrons. The goal is to measure the $ep$ differential cross sections in  the $Q^2$ range from 0.001 GeV$^2$  to  0.04 GeV$^2$   and to determine the proton charge radius with sub-percent precision.\\
       In the $ep$ elastic scattering experiments, the proton charge radius  is extracted from  the slope of  the electric form factor at the  momentum transfer squared $Q^2 \rightarrow $0.  In order to estimate the level of statistical and systematic errors in the extracted  proton radius, we simulated  the $ep$ elastic scattering differential cross section  using  the proton form factor available from analysis of  the   experimental data from the A1 experiment at Mainz. Then the proton radius was extracted from fitting the simulated pseudo-data with the cross section calculated using a $Q^2$ power series expansion of the proton electric form factor up to the $Q^8$ term. About 70 million of  the $ep$ elastic scattering events were generated in  the $Q^2$ range from  0.001 GeV$^2$ to 0.04 GeV$^2$,  that  corresponds to the  statistics  to be collected in our experiment in 45 days. For the considered $Q^2$ range and statistics, the main conclusions of these studies are as follows:
\begin{itemize}
\item  The extracted  value  of  the  proton  charge  radius  is  not  sensitive  to  the  $Q^8$  term,  so this  term  can  be neglected in the fits.  
\item  The fits with four free parameters ($A,<r_p^2>,<r_p^4>,<r_p^6>$)  determine  the proton charge $rms$-radius  $R_p$ = $<r_p^2>^{1/2}$  
with  the  errors  $\Delta R_p$   (stat)= 0.0085 fm (sigma)  and  $\Delta R_p (syst) \leq$ 0.001 fm.
\item  The statistical error  can be reduced  by a factor of  two  down to   $\Delta R_p (stat)$ = 0.0042 fm by fixing parameter  $<r_p^6>$  to some value determined in the experiments performed at larger transfer momenta. As an example, we  have  used  the  published value of  $<r_p^6>$  = 29.8 (7.6)(12.6) fm$^6$  determined in such experiments. Unfortunately, this value suffers from rather large systematic uncertainty that resulted in a systematic error  in the extracted proton radius : $\Delta R_p$ (syst)  = 0.0025 fm. Another promising approach is to use  a theoretical value for  $<r_p^6>$  in the fits .
\end{itemize}

\end{abstract}
\pagebreak

\section{Introduction}
\label{sec:Introduction}

The  striking difference  in the proton charge $rms$-radius extracted
from the two types of experiments,  the elastic $ep$ scattering experiments ( $R_p$ = 0.879 (5)(6) fm [1], Rp = 0.875(10) fm [2]) ) and  the muonic Lamb  shift experiments( $R_p$ = 0.8409 (4) fm [3] ),  so called ``proton radius puzzle'',  is widely discussed. As it is generally agreed,  new high precision measurements  of the $ep$ scattering  differential cross sections  in the low momentum transfer region are needed to resolve this puzzle. Recently, a new experiment was proposed by our collaboration [4]  to be carried out  in the 720 MeV electron beam of the Mainz Microtron MAMI. An innovative method will be used allowing for detection of recoil protons in coincidence with the scattered electrons. The goal of this experiment is to measure the $ep$ differential cross sections in the  $Q^2$ range from 0.001 GeV$^2$ to 0.04 GeV$^2$ with 0.1 \% relative and 0.2\% absolute precision and to determine the proton  charge radius with sub-percent precision. In this $Q^2$ range, about 70 million  $ep$ elastic scattering events should be collected in 45 days of the beam time.

   This note considers possible algorithms of analysis of the experimental data from this experiment.
In order to estimate the level of statistical and systematic errors in the extracted  proton radius, we simulated  the $ep$ elastic scattering differential cross section  using  the proton form factor  available  from analysis of  the experimental data from the A1 experiment at Mainz. Then the proton radius was extracted from fitting the simulated pseudo-data with the cross section calculated using various approximations for the $Q^2$ dependence of the proton form factor.



\section{Generation of $ep$ scattering events   }
\label{sec:Gen}
For this analysis, the $ep$ scattering events were generated according to  the following function describing the $ep$ elastic scattering differential cross section:  
 \begin{center}
 \vspace{-.4cm}
\begin{equation} 
\frac{d\sigma}{dt} =\frac{\pi \alpha ^{2} }{t^{2} } \left\{G_{E}^{2} \left[\frac{\left(4M+{t \mathord{\left/{\vphantom{t \varepsilon _{e}^{} }}\right.\kern-\nulldelimiterspace} \varepsilon _{e}^{} } \right)^{2} }{4M^{2} -t} +\frac{t}{\varepsilon _{e}^{2} } \right]-\frac{t}{4M^{2} } G_{M}^{2} \left[\frac{\left(4M+{t \mathord{\left/{\vphantom{t \varepsilon _{e} }}\right.\kern-\nulldelimiterspace} \varepsilon _{e} } \right)^{2} }{4M^{2} -t} -\frac{t}{\varepsilon _{e}^{2} } \right]\right\} 
\textnormal{GeV}^{-4}  \,,
\label{eq:Eq1} \,
 \vspace{.4cm}
\end{equation}
\end{center}
where  \textit{-t}  = $Q^2$; $\alpha$ = 1/137.036;   $M$ is the proton mass ($M$ = 938.272 MeV);   $\varepsilon_e$ is the total electron energy ($\varepsilon_e$ =720.5 MeV);  $G_E(Q^2$)  and  $G_M(Q^2$)   are  the electric and magnetic form factors, respectively.   We have accepted the following approximation valid for the small $Q^2$ region:
\begin{center}
\vspace{-.9cm}
\begin{equation} 
G_M(Q^2)  = \mu_p \cdot G_E(Q^2)  = 2.793 G_E(Q^2)  \,.
\label{eq:Eq2}
\end{equation}
\end{center}
$G_E(Q^2$)  is taken as a power series expansion:  
\begin{center}
\vspace{-.6cm}
\begin{equation} 
G_E(Q^2) = 1 - R2 \cdot B_2 \cdot Q^2 /C_2 + R4 \cdot B_4 \cdot  Q^4 /C_4 -  R6 \cdot B_6 \cdot  Q^6 /C_6 + R8 \cdot 
B_8 \cdot  Q^8 /C_8 \,\,,
\label{eq:Eq3} 
\end{equation}
 \end{center}
where  $B_n$ = $(5.06773)^n$,   $C_n$ = (n+1)!,  n=2,4,6,8;    $R2$ = $<r_p^2>$,  $R4$ = $<r_p^4>$ ,   $R6$ = $<r_p^6>$,  and $R8$ = $<r_p^8>$.  The \textit{rms}-radius $R_p$ = $(R2)^{1/2}$. In such presentation, $<r_p^n>$   and  $Q^n$ are  expressed in {fm$^n$}  and in GeV$^n$, respectively.  1 fm = 5.06773 GeV$^{-1}$;  1 GeV$^{-2}$ = 0.389379 mb.\\
The $ep$ scattering events were generated in the $Q^2$ range from   0.001 GeV$^2$ to 0.04 GeV$^2$ using              the values of $R2$, $R4$, $R6$, and $R8$ obtained by J.C.Bernauer [5,6]  from analysis of the cross sections measured in the A1 experiment:  
 \begin{center}
  $R2$ = 0.7700 fm$^2$,  $R4$ = 2.63 fm$^4$,  $R6$ = 26 {fm$^6$},    $R8$  = 374 fm$^8$ .
 \end{center}
The corresponding proton \textit{rms}-radius is  $R_p$ =$(R2)^{1/2}$ = (0.7700 fm$^2)^{1/2}$ = 0.8775 fm.\\
The $ep$ scattering cross sections integrated over the $Q^2$ range  $\,$ 0.001 GeV$^2 \leq Q^2 \leq$ 0.04 \, GeV $^2$    are:\\
$\sigma$($R_p$ = 0.8775 fm) = 0.248703 mb and $\sigma$($R_p$ = 0) = 0.254724 mb.\\
The ratio of these cross sections is $K$ = 0.976363.\\
As it follows from eqs.(\ref{eq:Eq1}) and (\ref{eq:Eq2}), the ratio of the differential cross sections gives the form factor squared in function of $Q^2$: 
\begin{center}
\vspace{-.9cm}
\begin{equation} \label{eq:Eq4}
d\sigma/dt (R_p = 0.8775 \textnormal{ fm} ) / d\sigma/dt (R_p = 0)  = G_E^2  (Q^2).                                               
\end{equation}
 \end{center}
We find this ratio by generating two similar samples of the $ep$
scattering events: one for $R_p$=0.8775 fm and another one for $R_p$ = 0. These samples should correspond to the same luminosity. That means that the number of generated events for  $R_p$ = 0.8775 fm should be by a factor  of $K$ = 0.976363 less than that for $R_p$ = 0. Then the  value of  $(G_E)_i^2$ in each bin can be obtained by the ratio of the numbers of generated events in that bin:  
\begin{center}
\vspace{-.8cm}
\begin{equation} 
(G_E)_i^2 = N_i ( R_p  = 0.8775 \textnormal{ fm} ) / N_i  ( R_p  = 0) \,.
\label{eq:Eq5} 
\end{equation}
 \end{center}
In order to reduce contribution of  fluctuations in $N_i$($R_p$=0) to the statistical error in $(G_E)_i^2$, the ($R_p$=0)  sample is  generated  with 100 times  larger  statistics, therefore  eq. (~\ref{eq:Eq5}) is transformed to:
\begin{center}
\vspace{-0.6cm}
\begin{equation} 
(G_E)_i^2   = N_i (R_p = 0.8775 \textnormal{ fm}) / 0.01 N_i(R_p = 0).      \label{eq:Eq6} 
\end{equation}
\end{center}
The  $ep$ scattering events  were generated  using the ROOT framework. Besides the analytical function of $d\sigma/dt$, we use as the input parameters: the $Q^2$ range,  the  binning within this range, and   the total number of generated events.  
At the level of the events generation, we use 1000 bins of equal width in the $Q^2$ range  0.001 GeV$^2 \leq Q^2 \leq 0.04 $ GeV$^2$ with  a possibility of further re-binning of the generated $G_E^2$ ($Q^2$) distribution. For each bin, the program gives the numbers of events integrated over the bin width,\\ $N_i$ ($R_p$ = 0.8775 fm)  and   $N_i$ ($R_p$ = 0),   and determines  $(G_E)_i^2$ according to eq.(~\ref{eq:Eq6}).  About 70 million events generated in the $Q^2$ range from  0.001 GeV$^2$  to   $Q^2$ = 0.04 GeV$^2$  correspond to the expected number of events to be  collected in our experiment in 45 days of continuous running with integrated luminosity $L_{int}$ = $2.8 \cdot 10^8$ mb$^{-1}$. As an example,  Figure 1 presents the simulated differential cross sections.  The total number of generated events was $N_{ev}$($R_p$ = 0.8775 fm) = $6.96369 \cdot 10^7$ events and   $N_{ev}$($R_p$ = 0) = 7.13227$\cdot 10^9$  events.  Figure 2 (left panel) shows the $G_E^2$ ($Q^2$) distribution determined according to eq.(~\ref{eq:Eq6}).  The right panel shows the same spectrum after re-binning the generated spectrum to 100 bins in  the same $Q^2$ range.
\vspace{-0.6cm}
\begin{SCfigure}[\sidecaptionrelwidth][hb]
\setlength{\unitlength}{1mm}
\begin{picture}(90,60)
    \put( 0,-5){\includegraphics*[width=90mm]{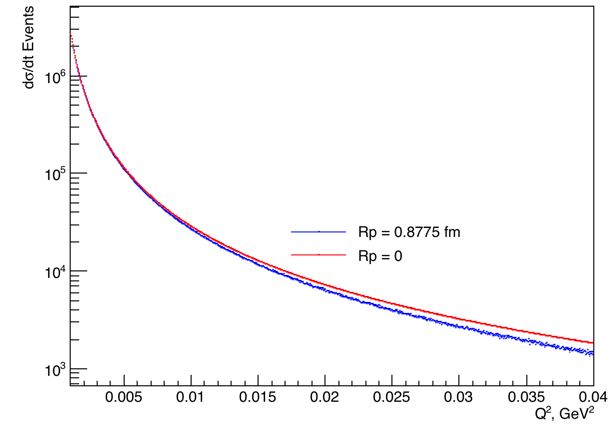}}
    \end{picture}
\caption {\small 
 Simulated differential cross \\sections for $R_p$=0.8775 fm (blue line) and for \\$R_p$ =0 (red line).
Statistics: \\$N_{ev}$($R_p$ =0.8775 fm) = $6.96369 \cdot 10^7$ events.\\
$N_{ev}$($R_p$=0) = 7.13227$\cdot 10^9$/100  events. \\
Binning: 1000 bins.   
  }    

\label{fig:Fig1}
\end{SCfigure}

\begin{figure}[ht]
  \setlength{\unitlength}{1mm}
  \centering
  \begin{picture}(180,60)
    \put( 0,0){\includegraphics*[width=90mm]{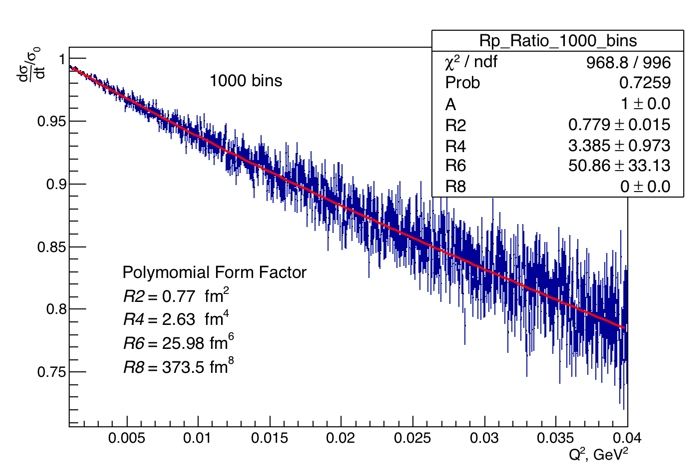}}
    \put(90,0){\includegraphics*[width=90mm]{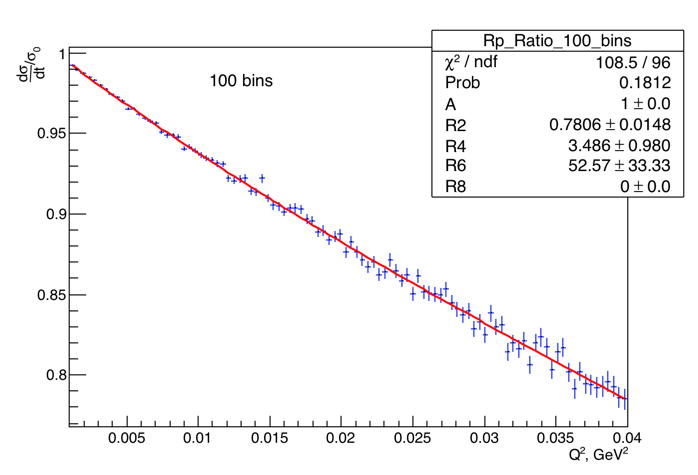}} 
   \end{picture}
  \caption {\small 
    Distribution  of  the  ratio  of  $d\sigma/dt$ ($R_p$= 0.8775 fm) / $d\sigma/dt$ ($R_p$= 0),  equivalent   to    the  $G_E^2$ ($Q^2$)  distribution,   obtained according to eq.(~\ref{eq:Eq6}).   \\Statistics:  $N_{ev}$($R_p$ = 0.8775 fm) = 6.9636 $\cdot 10^7$ events,  $N_{ev}$($R_p$ = 0) = 7. 13227$\cdot 10^9$  events.  Binning: 1000 bins (left panel)  and 100 bins (right panel).    Red lines  show  the results   of the fit with the form factor represented by Fit 1 in Table~\ref{tab:Tab1}. 
  }
\label{fig:Fig2}
\end{figure}
\section { Fitting of the $G_E^2$  ($Q^2$)  distributions}
\label{sec:Fit}
To fit the generated $G_E^2$ ($Q^2$) distributions, we use the power series expansion of the form factor: 
\vspace{-0.9cm}
\begin{center}
\begin{equation} 
G_E(Q^2)_{fit} = A \cdot (1 - R2 \cdot B_2 \cdot Q^2 /C_2 + R4 \cdot B_4 \cdot  Q^4 /C_4 -  R6 \cdot B_6 \cdot  Q^6 /C_6 + R8 \cdot 
B_8 \cdot  Q^8 /C_8)
\label{eq:Eq7} 
\end{equation}
\end{center}
\vspace{-0.3cm}
\noindent with the constants   $B_n$ and  $C_n$ as in eq.(~\ref{eq:Eq3}). The goal was to see how many $Q^2$ terms  should  be retained in this expression  to provide  minimal  combined  statistical  plus  systematic  error  in determination  of the proton radius.   The following options have been tested:\\
Option 1:  $A$, $R2$, $R4$, $R6$ are free parameters, $R8$ is a fixed variable.\\
Option 2:  $A$, $R2$, $R4$ are free parameters, $R6$ and $R8$ are fixed variables. \\

\textbf{Statistical errors in measurements of the proton radius} \\
\noindent Table~\ref{tab:Tab1} compares the statistical errors in $R2$ and $R4$ obtained   by fitting the generated $G_E^2$  ($Q^2$) with $G_E(Q^2)_{fit}$  represented by eq.(~\ref{eq:Eq7}) with four or three free parameters for statistics planned to collect in 45 days of continuous running of the experiment. 
\begin{table}[!hbt]
 \centering
\caption{\small 
 Comparison of statistical errors in $R2$ and $R4$  in the fits with three and four free parameters.\\ FF* denotes  parameters used to generate the $G_E^2$ ($Q^2$) distribution. Statistics: \\  $N_{ev}$($R_p$ = 0.8775 fm) = 6.9636 $\cdot 10^7$ events,  $N_{ev}$($R_p$ = 0) = 7. 13227$\cdot 10^9$  events.
 Binning: 1000 bins.  
  }
 \label{tab:Tab1}
 \vspace{-0.5cm}

 \begin{center}
 \begin{adjustbox}{width=0.8\textwidth , center}
\begin{tabular}{ | c | c | c | c | c | c |c |} 
\hline
& \makecell{$R2$, fm$^2$\\ $R_p$, fm } &  $R4$, fm$^4$  & $R6$, fm$^6$ & $R8$, fm$^8$ & $A$ & $\chi^2/ndf$ \\  
\hline
\textbf{FF*}   &\makecell{0.7700*\\ 0.8775} & 2.63* & 26* & 374* &  &  \\  \hline
\textbf{Fit 1} &\makecell{ 0.7790 (150) \\0.8826 (85)}  & 3.38 $\pm$ 0.97 & 51 $\pm$ 33       & \textbf{0} fixed& 1.0000(2)  & 969/996 \\ \hline
\textbf{Fit 2} &\makecell{ 0.7669 (72) \\ 0.8757 (41)}  & 2.52 $\pm$ 0.2  & \textbf{26} fixed & \textbf{0} fixed& 0.9999 (2) & 970/997 \\  \hline 
\end{tabular}
\end{adjustbox}
\end{center}
\end{table}
\newpage
\noindent From comparison of Fit 1 and Fit 2 in Table~\ref{tab:Tab1},  one can see that reduction of the number of free parameters
\textit{by fixing R6 to some fixed value reduces the statistical error in determination of the proton radius by a factor of two}
( from $\pm$ 0.0085 fm  to $\pm$ 0.0041 fm). Also, the $R4$ parameter is determined with 8\% precision in this fit. \\

\textbf{Systematic biases in measurement of the proton radius}\\

\noindent We have performed a number of fitting sets with various fixed values of $R6$ and $R8$ to study possible systematic biases related to this procedure. In each fitting set the fit was repeated 1000 times with independently generated $G_E^2$  ($Q^2$)  distributions. Figures 3 and 4 show the examples of such fits with four free parameters and with three free parameters, respectively.

\begin{figure}[h]
\center{\includegraphics[width=180mm]{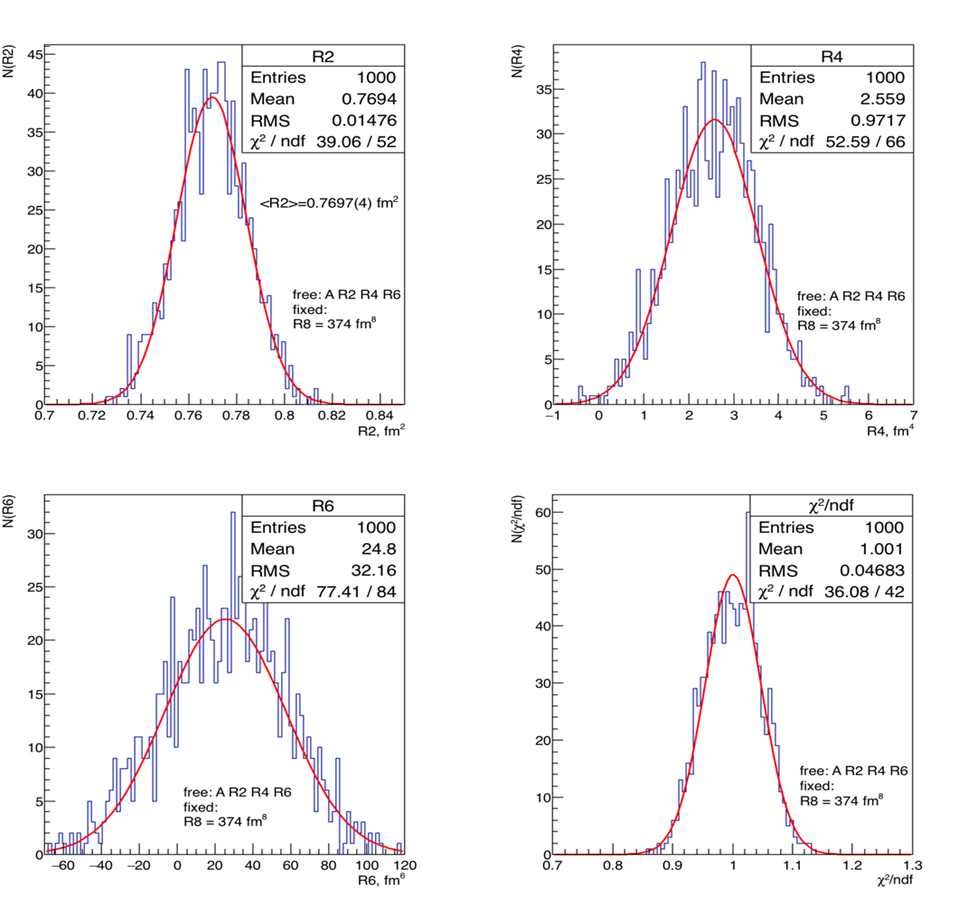}}
  \caption {\small 
  Distribution  of  the fitting   parameters  from  the fits  of   1000  independently  generated  $G_E^2$  ($Q^2$) distributions. The fitting function contained four free parameters $A, R2, R4, R6$ with   $R8$ = 374 fm$^8$. \\  Statistics: $N_{ev}$($R_p$ = 0.8775 fm) = 6.9636 $\cdot 10^7$ events in each $G_E^2$  ($Q^2$)  distribution. Binning: 1000 bins.  
  }
\label{fig:Fig3}
\end{figure}
\begin{figure}[h]
\center{\includegraphics[width=160mm]{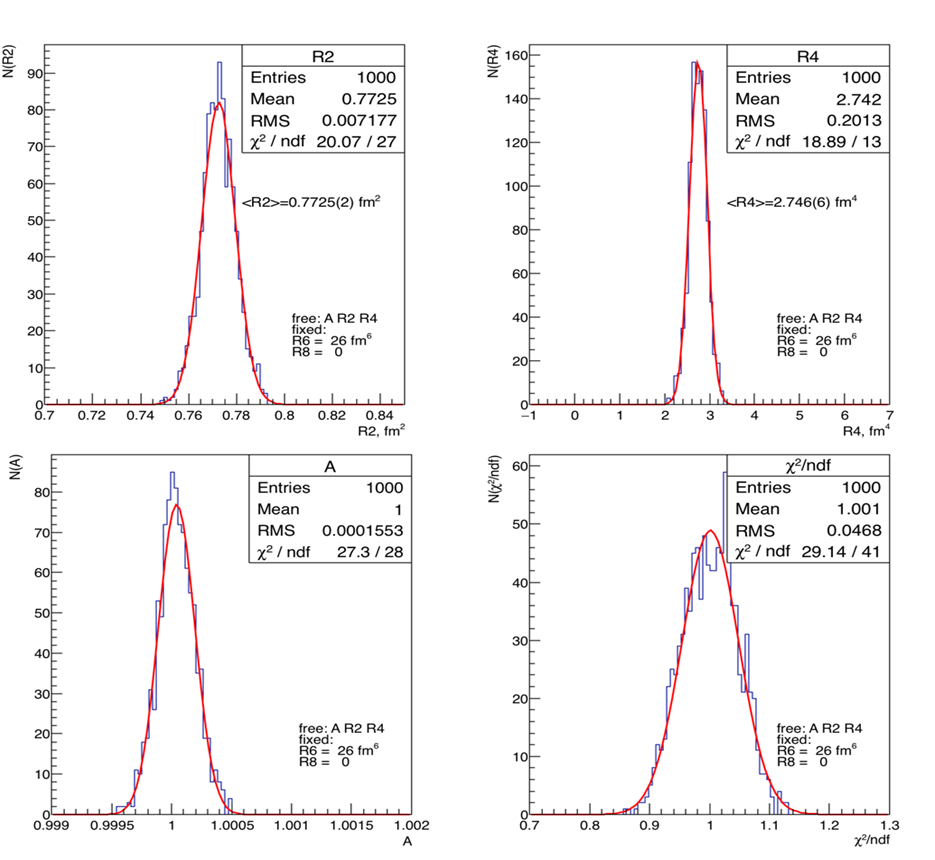}}
  \caption {\small 
Distribution of the fitting parameters obtained in the fits of 1000 independently generated  $G_E^2$  ($Q^2$)  distributions. The fitting function contained  three  free parameters $A$, $R2$, $R4$, with  fixed  $R6$ = 26 {fm$^6$} and  $R8$ = 0.
Statistics: $N_{ev}$($R_p$ = 0.8775 fm) = 6.9636 $\cdot 10^7$ events in  each    $G_E^2$  ($Q^2$)  distribution. Binning : 1000 bins.
   }
\label{fig:Fig4}
\end{figure}
\begin{figure}[h]
\center{\includegraphics[width=160mm]{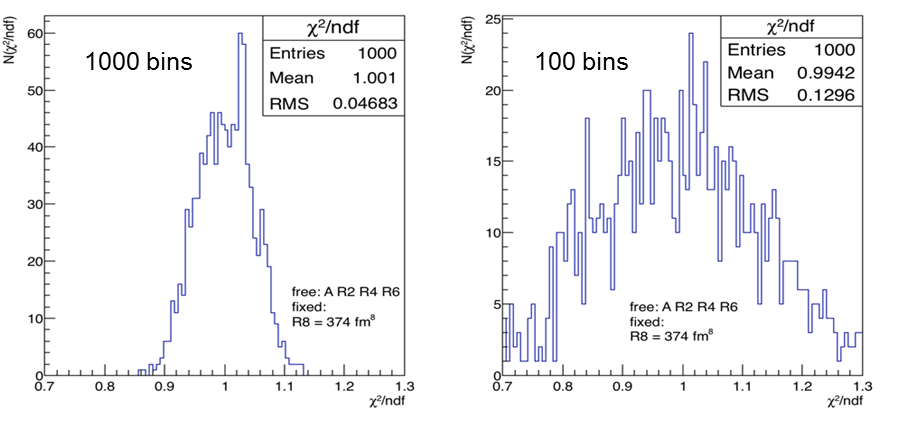}}
  \caption {\small 
Comparison of $\chi^2$/ndf   distributions obtained in fitting the same  $G_E^2$  ($Q^2$)  distributions subdivided in 1000 bins (left panel) and in 100 bins (right panel)    }
\label{fig:Fig5}
\end{figure}

\newpage
\noindent The distributions shown in Figs.~\ref{fig:Fig3},~\ref{fig:Fig4} were obtained with 1000 bins in the $G_E^2$ ($Q^2$) distributions.\\ The re-binning of these distributions to 100 bins gives identical fitting results, except the $\chi^2$ distribution  becomes  wider by  a factor of three (Fig.~\ref{fig:Fig5}). 

\begin{figure}[h]
\center{\includegraphics[width=180mm]{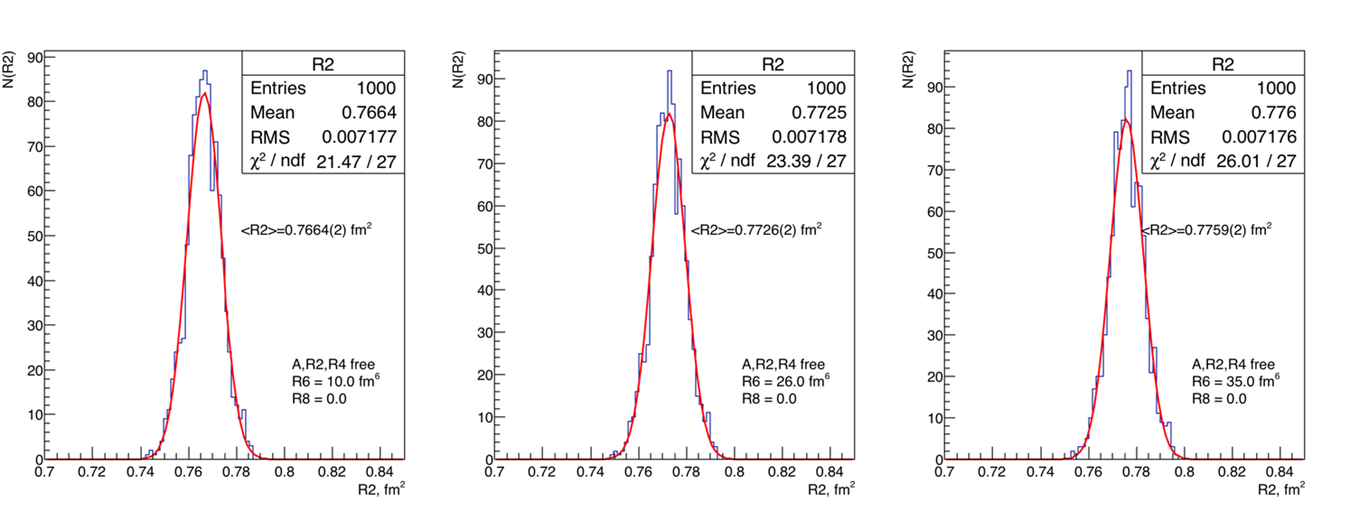}}
  \caption {\small 
Dependence  of  the $R2$ distributions  on variation  of the parameter  $R6$:  $R6$ = 10 {fm$^6$} (left panel),\\ 26 {fm$^6$} (central panel),  and  35 {fm$^6$}  (right panel).    $R8$ is set to zero.    Red lines -- fits  with  Gaussian distributions.  The width  of these distributions proved to be identical for all considered spectra. }
\label{fig:Fig6}
\end{figure}

\begin{figure}[!hbt]
\center{\includegraphics[width=180mm]{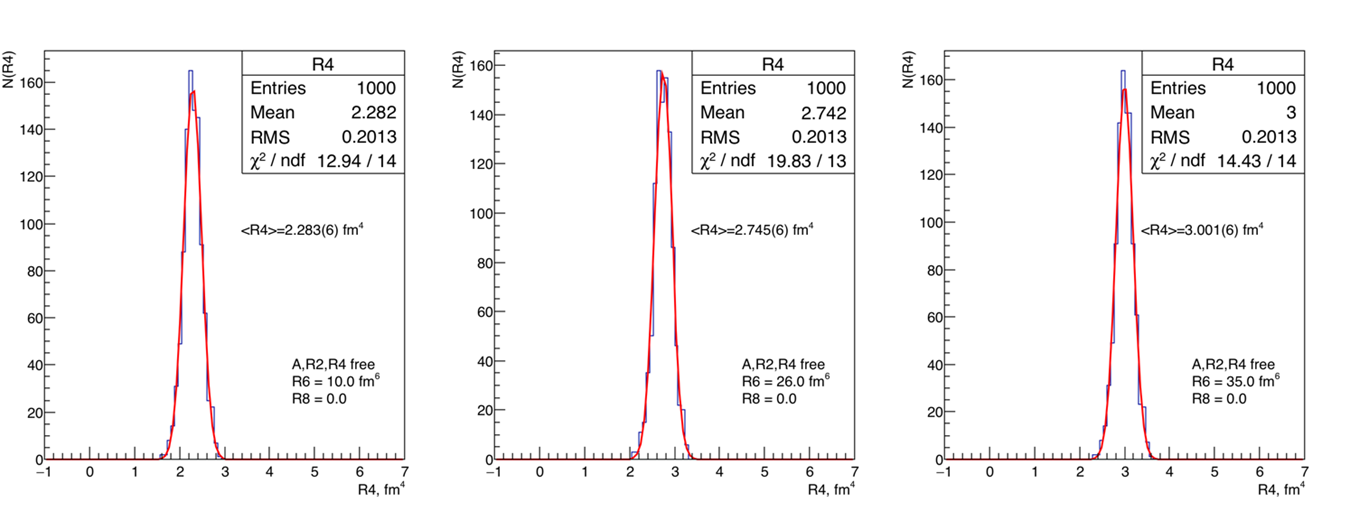}}
\caption {\small 
Dependence of the $R4$ distributions on variation of  the $R6$ parameter :  $R6$ = 10 {fm$^6$} (left panel),  26 {fm$^6$} (central panel),  and 35 {fm$^6$} (right panel).  $R8$ is set to zero. 
}
\label{fig:Fig7}
\end{figure}

As it follows from Fig.~\ref{fig:Fig4}, the fits with three free parameters can provide 0.0072/0.770 = 0.94\% statistical precision in determination of $R2$ (0.47\% precision in $R_p$).  In addition, $R4$ is measured with 8\% statistical precision. In these fits,  $R6$ and $R8$ were fixed to 26 {fm$^6$} and to zero, respectively. To see the sensitivity of obtained values of $R2$ and $R4$ to the chosen value of $R6$, the fits were repeated with $R6$ = 10 {fm$^6$} and 35 {fm$^6$}.  The results are presented in Figs.~\ref{fig:Fig6},~\ref{fig:Fig7} and in Table~\ref{tab:Tab2}.

As concerns the influence of parameter $R8$ on measurement of $R2$, it is proved to be  practically negligible, as it follows from comparison of Fit1 with Fit2 in Table~\ref{tab:Tab2}.  The variation of $R8$ from 374 fm$^8$ to zero shifts the value of $R2$ by 0.13\% (0.065\% shift in $R_p$).  On the other hand, the sensitivity of the extracted value of $R2$ to the fixed values of $R6$  is more essential   (Fits 3,4,5). The variation of $R6$ from 10 {fm$^6$} to 35 {fm$^6$} resulted in a systematic shift of $R2$ by 1.2 \% (0.6\% in $R_p$).
\pagebreak

\begin{table}[!hbt]
 \centering
\caption{\small 
 Mean  values   of  the  $R2$  and  $R4$  parameters   determined   from   the fits  of  1000  independently   generated\textbf{    }$G_E^2$  ($Q^2$)   distributions  (as shown in Fig.~\ref{fig:Fig2}) for  various  options  of  the fixed  parameters  $R6$  and  $R8$.  FF* denotes  the form factor parameters   used  to generate  the  $G_E^2$ ($Q^2$)  distribution.  In all shown fits,  the mean   values of parameters $A$ and $\chi^2$/ndf are equal to 1.0 with $10^{-4}$ and $10^{-3}$ accuracy, respectively.
  }
  \label{tab:Tab2}
   \vspace{-0.5cm}

 \begin{center}
\begin{tabular}{ | c | c | c | c | c |} 
\hline
\textbf{Fit } & \makecell{$<R2>$ \\ $<R2>-R2$* \\ fm$^2$} & \makecell{$<R4>$ \\ $<R4>-R4$* \\ fm$^4$} & \makecell{$<R6>$ \\ $<R6>-R6$* \\ {fm$^6$}} & \makecell{$R8$ \\  fm$^8$} \\ \hline 

\makecell{\textbf{FF*}\\} & 0.7700* & 2.63* & 25.98* & 373.5* \\ \hline 
\textbf{Fit1} &\makecell{ 0.7703(5)\\ +0.0003(5) }&\makecell{ 2.61$\pm$0.03\\ -0.03(3) } &\makecell{ 26$\pm$1\\ 0$\pm$1 } &\makecell{ \textbf{374}\\ fixed }\\ \hline 
\textbf{Fit2} &\makecell{ 0.7693(5)\\ -0.0007(5) } &\makecell{ 2.49$\pm$0.03\\ -0.14(3) } &\makecell{ 23.4$\pm$1.0\\ -1.6$\pm$1.0 } &\makecell{ \textbf{0}\\ fixed} \\ \hline 
\textbf{Fit3} &\makecell{ 0.7727(2)\\ +0.0027(2) } &\makecell{ 2.743(7)\\ +0.113(7) } &\makecell{ \textbf{26}\\ fixed } &\makecell{ \textbf{0}\\ fixed } \\ \hline 
\textbf{Fit4} &\makecell{ 0.7665(2)\\ -0.0035(2) } &\makecell{ 2.284(7)\\ -0.346(7) } &\makecell{ \textbf{10}\\ fixed } &\makecell{ \textbf{0}\\ fixed } \\ \hline 
\textbf{Fit5} &\makecell{ 0.7761(2)\\ +0.0061(2) } &\makecell{ 3.00(7)\\ +0.37 } &\makecell{ \textbf{35}\\ fixed  }&\makecell{  \textbf{0}\\ fixed}\\ \hline 

\end{tabular}
\end{center}
\end{table}

The systematic biases were studied also by another method when the simulated cross sections were generated with 1000 times higher statistics: ($N_{ev}$($R_p$ = 0.8775 fm) = 6.96369$\cdot 10^{10}$  events and \\ $N_{ev}$ ($R_p$ = 0) = 7.13227$\cdot 10^{10}$  events). The results are presented in Table~\ref{tab:Tab3}. As it follows from Fits 1,2,3 in Table~\ref{tab:Tab3}, variation  of  $R8$  from  $R8$ = 0  to $R8$ =700 fm$^8$  resulted in a  0.2\% shift in the extracted  $R2$ value.  Therefore, it is safe to fix $R8$   at $R8$ = 374 fm$^8$ and consider the systematic error in $R2$ due to uncertainties in $R8$ to be on a level of $\pm$ 0.1 \%  ( 0.05\% in $R_p$).   

While fixing the $R6$ parameter, it is natural to take into account the results of previous analyses of the $ep$ scattering data. According to [6],  $R6$ = 29.8 (7.6)(12.6) {fm$^6$} and $R4$  = 2.59 (19)(04) fm$^4$ . Therefore, we can fix $R6$ at $\simeq$ 26 {fm$^6$} with uncertainty of  $\pm$ 15 {fm$^6$}. As one can see from  Fits 4, 5, 6  in Table~\ref{tab:Tab3},   such uncertainty in $R6$  leads to  $\pm$ 0.8\% systematic errors in  $R2$ ($\pm$ 0.4\% in $R_p$).
As to the $R4$ parameter, it can be determined directly from our experimental data, and comparison with the A1 data could be used as a cross check.\\
\\

Additional   study of  the systematic shifts in the $R2$ values was done by  fitting    the ratio of        the differential cross sections $d\sigma/dt$($R_p$) / $d\sigma/dt$($R_p$ =0)  generated with high statistics for  three options of the polynomial Form Factor,  FF1, FF2, and FF3,  with variations of the $R4$, $R6$, and $R8$ values  consistent with the uncertainties  of the A1 data. The fitting function contained three free parameters ($A$, $R2$, $R4$), while the $R6$ and $R8$ parameters were fixed to 26 {fm$^6$} and to 374 fm$^8$, respectively. The results are presented in Table~\ref{tab:Tab4}.

\begin{table}[h]
 \centering
\caption{\small 
The results  of  fitting  of  the  $G_E^2 $($Q^2 $) distribution  obtained  from the  ratio of the differential cross sections $d\sigma $/$dt$($R_p$ = 0.8775 fm) / $d\sigma $/$dt$($R_p$ =0)  generated with high statistics  ($N_{ev} (R_p$ = 0.8775 fm) = 6.96369$\cdot 10^{10}$events, $N_{ev} (R_p$ = 0)  = 7.13227$\cdot 10^{10}$events).  A polynomial form factor FF* was used to generate  $d\sigma $/$dt$($R_p$ = 0.8775 fm) with  $R2$, $R4$, $R6$, and $R8$ parameters (denoted by FF*)  taken from the analysis of the A1 data [5].  The generated pseudo-data were fitted with a polynomial function with various options of the fixed $R8$ and $R6$ parameters.
  }
  \label{tab:Tab3}
\vspace{-0.5cm}
 \begin{center}
\begin{adjustbox}{width=0.9\textwidth , center}
\begin{tabular}{ | c | c | c | c | c |c | c | c |} 
\hline
&  {$R2 -R2$* ,  fm$^2$} & {$R2$ ,  fm$^2$} & {$R4$ ,  fm$^4$} & { $R6$ ,  {fm$^6$}} & {$R8$, fm$^8$} & \textbf{$\chi^2/ndf$} & \textbf{$A$} \\ \hline 
\makecell{FF*\\} &  & 0.7700* & 2.63* & 26* & 374* &  &  \\ \hline 
\makecell{FIT1\\} & --0.0013 (6) & 0.7687 (6) & 2.54 (4) & 22.9 (1.4) & \textbf{374} fixed & 963/996 & 1.00000 (1) \\ \hline 
\makecell{FIT2\\} & --0.0022 (6) & 0.7678 (6) & 2.43 (4) & 13.9 (1.4) & \textbf{0} fixed & 966/996 & 1.00000 (1) \\ \hline 
\makecell{FIT3\\} & --0.0002 (3) & 0,7698 (3) & 2.64 (4) & 30.9 (1.4) & \textbf{0} fixed & 966/996 & 1.00000 (1)\\ \hline 
\makecell{FIT4\\} & --0.0002 (3) & 0.7698 (3) & 2.63 (1) & \textbf{26} fixed & \textbf{374} fixed & 968/997 & 1.00000 (1) \\ \hline 
\makecell{FIT5\\} & --0.0064 (3) & 0.7636 (3) & 2.16 (1) & \textbf{10} fixed & \textbf{374} fixed & 1046/997 & 1.00000 (1) \\ \hline 
\makecell{FIT6\\} & +0.0054 (3) & 0.7753 (3) & 3.03 (1) & \textbf{40} fixed & \textbf{374} fixed & 1106/997 & 1.00000 (1) \\ \hline 
\end{tabular}
\end{adjustbox}
\end{center}
\end{table}
\pagebreak
\noindent Table~\ref{tab:Tab4} shows that the fits with a fixed $R6$ parameter ($R6$ =26 {fm$^6$}) reproduce $R2$ with    $\pm$ 0.56 \% systematic error ($\pm$ 0.28\% error in the proton radius), assuming that the $R6$ value in the real experimental data will be in the limits 11 {fm$^6$} $< R6 <$ 41 {fm$^6$}.  

\begin{table}[!hbt]
 \centering
\caption{\small 
The results of fitting  of  the $G_E^2$($Q^2$) distributions obtained from the  ratio of the differential cross sections $d\sigma/dt$($R_p$ = 0.8775 fm) / $d\sigma/dt$($R_p$ =0)  generated with high statistics ($N_{ev}$ ($R_p$ = 0.8775 fm) = 6.96369$\cdot 10^{10}$  events, $N_{ev}$ ($R_p$ = 0)  = 7.13227$\cdot 10^{10}$  events) with three options  of the polynomial form factor, FF*1, FF*2, FF*3,  consistent with the uncertainties of the A1 data. The generated pseudo-data were  fitted with a polynomial function
$G_E(Q^2) = A \cdot (1 - R2 \cdot B_2 \cdot Q^2 /C_2 + R4 \cdot B_4 \cdot  Q^4 /C_4 -  R6 \cdot B_6 \cdot  Q^6 /C_6 + R8 \cdot 
B_8 \cdot  Q^8 /C_8)$
 with   the $R6$ and $R8$ parameters fixed to 26 {fm$^6$} and to 374 fm$^8$, respectively. The $A$ parameter  proved  to be 1.00000 with $10^{-5}$ error in all  fits.
  }
\label{tab:Tab4}
 \vspace{-0.5cm}

\begin{center}
\begin{adjustbox}{width=0.9\textwidth , center}
\begin{tabular}{ | c | c | c | c | c |c | c |} 
\hline
\textbf{Fit\#} & $R2$, fm$^2$ & ($R2-R2$*), fm$^2$ & $R4$, fm$^4$ & $R6$, fm$^6$ & $R8$, fm$^8$ & \textit{$\chi^2/ndf$} \\ 
\hline 
\makecell{FF*1\\ Fit} &\makecell{ 0.7700*\\ 0.7699(3)} & --0.0001(3) &\makecell{  2.63*\\ 2.626(3) }&\makecell{  26.0*\\ \textbf{26} fixed }&\makecell{ 374*\\ \textbf{374} fixed} & 968/997 \\ \hline 
\makecell{FF*2\\ Fit} &\makecell{ 0.7700*\\ 0.7742(3) } & +0.0042(3) &\makecell{ 2.43*\\ 2.772(9)  }&\makecell{ 11*\\ \textbf{26} fixed  }&\makecell{ 160*\\ \textbf{374} fixed} & 1052/997 \\ \hline 
\makecell{FF*3\\ Fit} &\makecell{ 0.7700*\\ 0.7656(3)} & --0.0044(3) &\makecell{ 2.83*\\ 2.482(3) }&\makecell{ 41*\\ \textbf{26} fixed  }&\makecell{ 600*\\ \textbf{374} fixed } & 988/997 \\ \hline 
\end{tabular}
\end{adjustbox}
\end{center}
\end{table}

\newpage
\section {Summary  }
\label{sec:Summ}
We have analyzed the simulated  $ep$ scattering differential cross section  expected from an experiment aimed at high precision measurement of the proton charge \textit{rms}-radius  $R_p$ = $<r_p^2>^{1/2}$ . Following   the Proposal of our experiment, it was accepted that 70 million of the $ep$ elastic scattering events will be collected in the $Q^2$ range  0.001 GeV$^2 \leq Q^2 \leq $ 0.04 GeV$^2$.    The $ep$ elastic scattering events were generated with  the polynomial proton charge form factor determined by J.C.Bernauer et al. in the data analysis of the A1 experiment [5,6], with an additional assumption that $G_M$($Q^2$) = $\mu_p \cdot G_E(Q^2$) in the considered $Q^2$ range. The generated pseudo-data were fitted with a polynomial function: \\
$G_E(Q^2) = A \cdot (1 - <r_p^2> \cdot B_2 \cdot Q^2 /C_2 + <r_p^4> \cdot B_4 \cdot  Q^4 /C_4 -  <r_p^6> \cdot B_6 \cdot  Q^6 /C_6 + <r_p^8> \cdot 
B_8 \cdot  Q^8 /C_8)$ ,   
where  $B_n$ = (5.06773)$^n$,   $C_n$ = (n+1)!, n = 2,4,6,8;  $<r_p^n>$ and $Q^n$ are  expressed in {fm$^n$}  and in GeV$^n$, respectively. Two options have been tested:

Option 1:  $A$, $<r_p^2>$, $<r_p^4>$, $<r_p^6>$  are free parameters,  $<r_p^8>$ is  a fixed  variable;

Option 2:  $A$, $<r_p^2>$, $<r_p^4>$ are free parameters,  $<r_p^6>$ and $<r_p^8>$ are fixed variables. \\

The results of the analysis can be summarized as follows:\\
\begin{itemize}
\item 
  The $Q^8$ term plays very little role in determination of  $R_p$. The variation of  $<r_p^8>$ from zero to 700 fm$^8$  leads to increasing the $R_p$ value by 0.001 fm. Therefore, one can fix $<r_p^8>$, for example,  at the value  from the A1 analysis ( $<r_p^8>$ = 374 fm$^8$ [5]). This may introduce a systematic error in $R_p$ due to uncertainties in $<r_p^8>$  on a negligible level of  $\pm$ 0.0005 fm.
\item 
   The  statistical  error   in  $R_p$  in   the  fits  with  four   free  parameters ($A$,  $<r_p^2>$,  $<r_p^4>$,  $<r_p^6>$)\\             
is $\pm$ 0.0085 fm. The advantage of such fit is a negligibly small systematic bias.
\item
  The statistical error in $R_p$ can be reduced by a factor of two (down to $\pm$ 0.0042 fm) in the fit with three free parameters ($A$, $<r_p^2>$, $<r_p^4>$) by fixing  $<r_p^6>$  to some value followed from the analysis of the $ep$ scattering data in the higher $Q^2$ region. However, in this case some systematic bias may be introduced because of uncertainties in the $<r_p^6>$  value. The sensitivity of $R_p$ to variations in $<r_p^6>$, as determined in our analysis, is  as follows: a shift in $<r_p^6>$ by  6 {fm$^6$} produces a shift in $R_p$ by 0.001 fm.  
\item
  The existing polynomial  fits to the available $ep$ scattering data  determined various moments of  the proton form factor $<r_p^n>$  [5,6]. In particular, it was found that $<r_p^6>$ = 29.8 (7.6)(12.6) {fm$^6$}.  Unfortunately, this result suffers from a large systematic error, which corresponds to \\ a  $\pm$ 0.0025 fm systematic bias in the extracted value of the proton radius $R_p$.
\item
   Another approach to the proton form factor was demonstrated recently  by J.$M$. Alarcon et al. [7,8].  On the basis of the Dispersive Improved Chiral Effective Field Theory, they calculated various  FF moments from $<r_p^2>$ to $<r_p^{20}>$ with remarkably small error bars.  Their predictions for the lowest moments of the charge FF are:   $<r_p^2>$ = (0.701, 0.768) fm$^2$,   $<r_p^4>$ = (1.47, 1.6) fm$^4$, $<r_p^6>$ = (8.5, 9.0) {fm$^6$},  $<r_p^8>$ = (127, 130) fm$^8$. Note that precision of the calculations is higher for higher FF moments in this approach, so it looks safe to take the predicted values of  $<r_p^6>$ = 9.0 {fm$^6$}  and $<r_p^8>$ = 130 fm$^8$  for our fits. The systematic bias will be negligible in this case, even  assuming the real error in  $<r_p^6>$ will be an order of magnitude larger than that quoted above. 
\item
   Besides the proton radius $R_p$,  the $<r_p^4>$  parameter  will  be also determined with 8\% statistical errors in the  fits with fixed $<r_p^6>$ and $<r_p^8>$ .
\end{itemize}
\pagebreak
In conclusion, Table~\ref{tab:Tab5} presents the statistical and systematic errors  related  to the procedure of extraction of the proton charge radius from the experimental data expected  in our experiment.

\begin{table}[h]
 \centering
\caption{\small 
Statistical    and    systematic  errors   in   $R_p$  resulted  in   the  fits  of  the  psuedo-data  with  a  polynomial  function 
$G_E(Q^2) = A \cdot (1 - <r_p^2> \cdot B_2 \cdot Q^2 /C_2 + <r_p^4> \cdot B_4 \cdot  Q^4 /C_4 -  <r_p^6> \cdot B_6 \cdot  Q^6 /C_6 + <r_p^8> \cdot B_8 \cdot  Q^8 /C_8)$  
with  three  or  four    free parameters.   
Statistics : 7 $\cdot 10^7 ep$ scattering events in the $Q^2$ range 0.001 GeV$^2 \leq Q^2 \leq $ 0.04 GeV $^2$.    
  }
 \label{tab:Tab5}
 \vspace{-0.5cm}
 \begin{center}
 \begin{adjustbox}{width=0.8\textwidth , center}

\begin{tabular}{ | c | c | c | c | c | c |} 
\hline
 &\makecell{Free \\parameters}   &  \makecell{Fixed\\ parameters} & \makecell{$\Delta$ $R_p$ (stat) } & \makecell{$\Delta$ $R_p$ (syst) } & \makecell{comments} \\ \hline 

\textbf{Option1} & \makecell{$A$  $<r_p^2>$  $<r_p^4>$\\ $<r_p^6>$ } & $<r_p^8>$ & $\pm$ 0.0085 fm  & $<$ 0.001fm & ---- \\ \hline
\textbf{Option2} &\makecell{ $A$ $<r_p^2> <r_p^4>$}  & $<r_p^6> <r_p^8>$ & $\pm$ 0.0042 fm  &\makecell{ $\pm$ 0.0025 fm \\  $<$ 0.001fm  } & \makecell{$<r_p^6>$ from[6]\\ $<r_p^6>$ from[7]} \\ \hline

\end{tabular}
\end{adjustbox}
\end{center}
\end{table}

 Some other options of the analysis are presented in the ANNEXes to this note.

\section*{References}

\begin{enumerate}
 \item 
 J. Bernauer et al., \textit{High-Precision Determination of the Electric and Magnetic Form Factors of the Proton}, \,\, A1 Collaboration, Phys. Rev. Lett. 105, 242001 (2010).
 \item 
 X. Zhan et al., \,\,\textit{High-precision measurement of the proton elastic form factor ratio at low $Q^2$} Phys.Lett.B 705, 59, (2011).
 \item 
 A. Antognini et al., \textit{Proton Structure from the Measurement of 2S-2P Transition Frequencies of Muonic Hydrogen}, \,\, Science  339, 417 (2013).
 \item 
 A. Vorobyev,  A. Denig,  \textit{High Precision Measurement of the ep elastic cross section at small  $Q^2$}, Proposal to perform an experiment at the A2 hall, MAMI, November  2017.
 \item 
 J. C. Bernauer, Ph.D. thesis, University of Mainz, 2010.
 \item 
 M. O. Distler, J. C. Bernauer, and T.Walcher, \textit{The RMS Charge Radius of the Proton and Zemach Moment},   arXiv:1011.1861 v3[nucl-th], June 2011.
 \item 
 J.M. Alarcon, C.Weiss, \textit{Accurate nucleon electromagnetic form factors from dispersively improved chiral effective field theory},  ArXiv: 1803.09748 [hep-ph]  Phys.Lett.B 784 (2018) 373.
 \item 
 J.M. Alarcon et al., \textit{Proton charge radius extraction from electron scattering data using dispersively improved chiral effective field theory}, ArXiv: 1809.06373 [hep-ph].  

\end{enumerate}

\pagebreak

\section*{Annex 1. Fits with fixed ratio  $\eta= R6/R4$  } 

The parameter $R6$  is rather strongly correlated with $R4$ as it can be  seen from  Table \ref{tabA1:TabA1-1}.


\begin{table}[h]
 \centering

\caption{\small 
The  values  of   R4 and  R6  in  different    presentations  of the proton Form Factor,
corresponding to\\ R2 = 0.7700 fm$^2$
  }

\begin{center}
\begin{tabular}{ | c | c | c | c |} 
\hline
Form Factor & $R4$,  fm$^4$ & $R6$,  {fm$^6$}& $\eta = R6 / R4$,  fm$^2$ \\  \hline 
Dipole FF          &   1.49 &    5.3 & 3.6 \\ \hline 
Di$\chi $EFT [7,8] &   1.6  &    9.0 & 5.6 \\ \hline 
Bernauer [5]       &   2.63 &    26  & 9.9 \\ \hline 
\end{tabular}
\label{tabA1:TabA1-1}
\end{center}
\end{table}

\begin{figure}[!htb]
  \setlength{\unitlength}{1mm}
  \centering
  \begin{picture}(160,190)
    \put( 030,0){\includegraphics*[width=95mm]{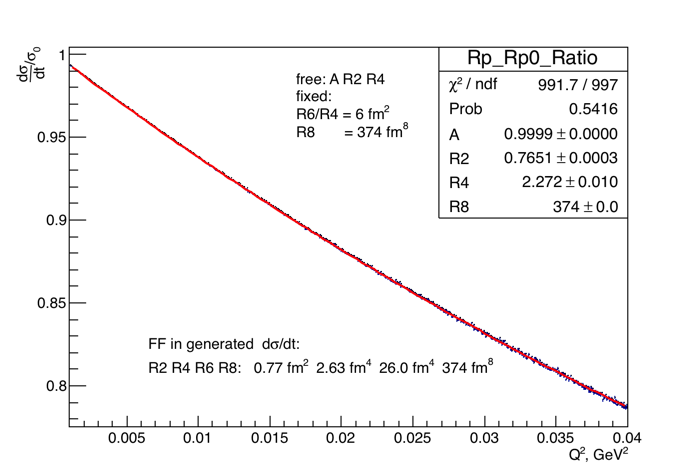}}
    \put( 0,60){\includegraphics*[width=80mm]{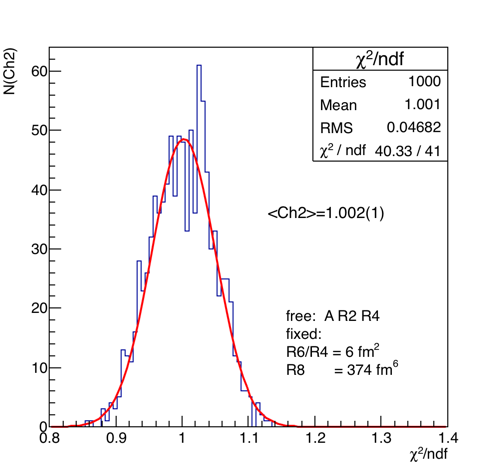}}
    \put(75,60){\includegraphics*[width=80mm]{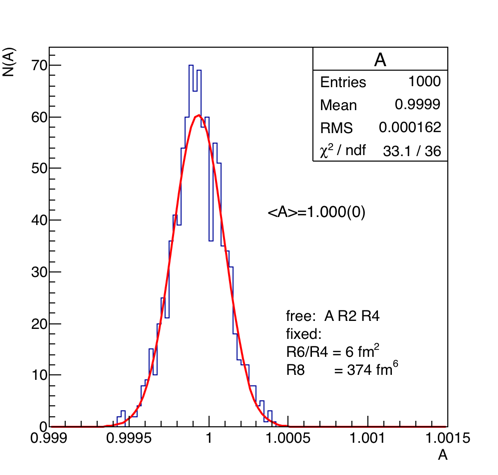}} 
     \put( 0,130){\includegraphics*[width=80mm]{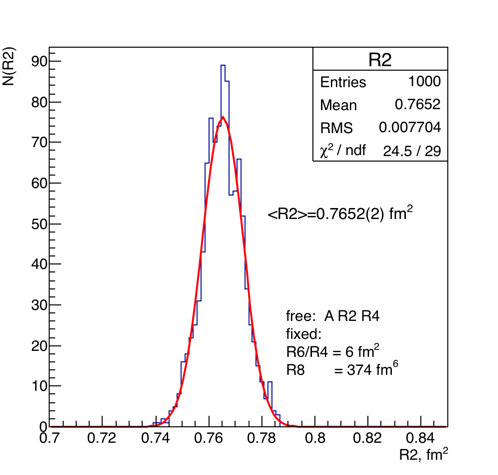}}
    \put(75,130){\includegraphics*[width=80mm]{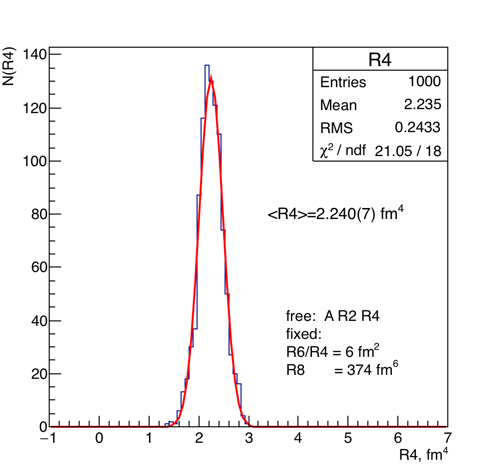}} 
    \put( 55,50){e)}
    \put( 15,120){c)}
    \put( 90,120){d)}
    \put( 15,190){a)}
    \put( 90,190){b)}
  \end{picture}
\caption {\small 
\\
 \textbf{Panels  a)  b)  c)  d).}  Distribution of the fit parameters obtained from 1000 independent fits  of $d\sigma/dt$ generated  in the  $Q^2$ range  from 0.001 GeV$^2$ to 0.04 GeV$^2$  using Bernauer's proton form factor. \textit{Statistics}:  7$\cdot 10^7$ events in each generated set. \textit{Binning} : 1000 bins.  Fitting with binomial FF containing up to $Q^8$ term.  \textit{Free parameters}: $A$, $R2$,$R4$. 
\textit{Fixed parameters:} $R8$ = 374 fm$^8$  and  \textbf{ $R6$/$R4$ = 6 fm$^2$.}\\
\textbf{Panel  e). }Results of one fitting  set with super high statistics :  7~$\cdot 10^9$  events. All fit conditions are as above. 
 } 
\label{fig:FigA-1}
\end{figure}

Therefore, instead of $R6$, one can try to use in the fitting function the ratio $\eta = R6/R4$. That is, instead of eq.(~\ref{eq:Eq7}), to use the following expression in the fits:
\vspace{-0.9cm}
\begin{center}
\begin{equation} 
G_E(Q^2)_\textit{fit} = A \cdot (1 - R2 \cdot B_2 \cdot Q^2 /C_2 + R4 \cdot B_4 \cdot  Q^4 /C_4 - \eta \cdot R_4 \cdot B_6 \cdot  Q^6 /C_6 + R8 \cdot B_8 \cdot  Q^8 /C_8) \,,
\label{eq:A-Eq7} 
\end{equation}
\end{center}
\vspace{-0.3cm}
where $\eta$  is a variable parameter.   This fitting function was used to fit the pseudo-data generated with the Bernauer's Form Factor, following the procedure described above in this note.  The value of $\eta$ was varied from $\eta$= 6  to $\eta$= 12, with $R8$ = 374 fm$^8$.   The fitting procedure is illustrated  by Fig.~\ref{fig:FigA-1} which shows the distribution of the fit parameters $A$, $R2$, $R4$, and $\chi^2$/ndf obtained in the fits with the regular statistics (panels a), b), c), d)). Also, this Figure (panel e))  shows an example of the super high statistics fit used for studies of the systematic shifts in the measured values of $R2$ and $R4$ in dependence on the value of the ratio $\eta$.  The results of these studies are presented in Table~\ref{tabA1:TabA1-2}.

\begin{table}[h]
 \centering
\caption{\small 
The results  of  fitting   the  ratio  of  the  differential  cross sections   $d\sigma/dt$($R_p$ = 0.8775 fm) / $d\sigma/dt$($R_p$ =0)   generated with   the  Bernauer's  Form  Factor FF*. The generated  pseudo - data  were   fitted  with  a  polynomial function $G_E(Q^2)  = A \cdot (1 - R2 \cdot B_2 \cdot Q^2 /C_2 + R4 \cdot B_4 \cdot  Q^4 /C_4 - \eta \cdot R_4 \cdot B_6 \cdot  Q^6 /C_6 + R8 \cdot B_8 \cdot  Q^8 /C_8) $ 
($R_p$ = 0.8775 fm) = 6.96369$\cdot 10^{10}$  events.      Binning: 1000 bins
  }

  \label{tabA1:TabA1-2}
\vspace{ -0.4cm}
\begin{center}
\begin{adjustbox}{width=0.9\textwidth , center}
\begin{tabular}{ | c | c | c | c | c |c | c |} 
\hline
Fit\# & $R2$, fm$^2$ & $R2$ -- $R2$* , fm$^2$ & $R4$ , fm$^4$ & $R6$/$R4$,  fm$^2$ & $R8$, fm$^8$ & $\chi^2/ndf$ \\ \hline 
\makecell{FF*  } & 0.7700*   &                & 2.63*   & 9.9*     & 374*       & \\ \hline 
\makecell{Fit 1 }& 0.7651(3) &   -- 0.0049(3) & 1.33(1) & \textbf{6}  fixed & \textbf{374}  fixed &   992/997 \\ \hline 
\makecell{Fit 2 }& 0.7674(3) &   -- 0.0026(3) & 1.48(1) & \textbf{8}  fixed & \textbf{374} fixed  &   975/997 \\ \hline 
\makecell{Fit 3 }& 0.7700(3) &      0.0000(3) & 1.62(1) & \textbf{10} fixed & \textbf{374} fixed  &   953/997 \\ \hline 
\makecell{Fit 4 }& 0.7729(3) &    + 0.0029(3) & 1.91(1) & \textbf{12} fixed & \textbf{374} fixed  &  1010/997 \\ \hline 
\end{tabular}
\end{adjustbox}
\end{center}
\end{table}

As it follows from Table~\ref{tabA1:TabA1-2}, the variation of the ratio $R6/R4$  from 6 fm$^2$ to 12 fm$^2$ resulted in              a  1\% shift in the value of $R2$  ( 0.5\% shift in $R_p$).\\
In other words,   with  the ratio  $R6$/$R4$   fixed  to  8 fm$^2$, one can expect a systematic bias in  the  measured \textit{rms}-proton radius  $\Delta R_p = \pm$ 0.0014 fm, assuming that in the real experimental data  this ratio   will be between  6 fm$^2$ ( Di$\chi $EFT)  and  10 fm$^2$ (Bernauer).

\pagebreak

\pagebreak

\section*{Annex 2. Dipole Form Factor  } 

\begin{figure}[h]
\setlength{\unitlength}{1mm}
\centering
\begin{picture}(160,40)
     \put( 0,0){\includegraphics*[width=160mm]{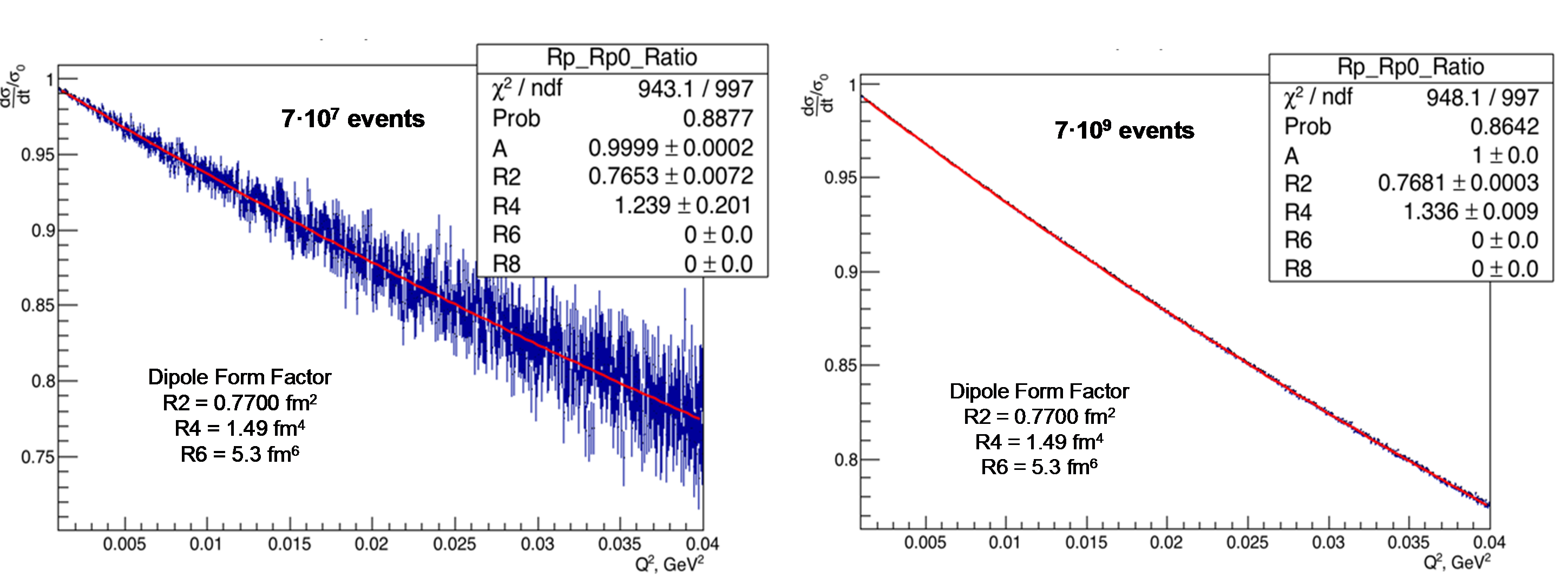}}
\end{picture}
\caption {\small
Distribution of the  ratio  $d\sigma/dt (R_p$ = 0.8775 fm) $/d\sigma/dt (R_p=0)$   generated with a modified Dipole Form Factor. Statistics: $N_{ev} (R_p$ = 0.8775 fm) = 6.9636 $\cdot 10^7$ events (left panel),  
$N_{ev}(R_p$ = 0.8775 fm) = 6. 96369 $\cdot 10^9$  events (right panel). Binning: 1000 bins. Red lines show the results of the fit with the form factor represented by Fit 1 in Table \ref{tabA2:TabA2-1}
}
\label{fig:FigA2-1}
\end{figure}

Similar analysis was performed using a modified Dipole Form Factor in the generated differential cross section $d\sigma/dt$ :
\vspace{-.5cm}
 \begin{center}
$G_E(Q^2)$  = $(1 + Q^2 /0.6068)^{-2}$. 
\end{center}
\vspace{-.3cm}
The  power series expansion of this form factor corresponds to the following parameters:
\vspace{-.2cm}
\begin{center}
$<r_p^2>$ = 0.7700 fm$^2$ ,  $<r_p^4>$ = 1.49 fm$^4$ ,   $<r_p^6>$ =  5.3 {fm$^6$} , $R_p$ =  $<r_p^2>^{1/2}$ =  0.8775 fm.  
\end{center}
\vspace{-.3cm}
The cross sections  integrated over the $Q^2$ range 0.001 GeV$^2 \leq Q^2 \leq $ 0.04 GeV$^2$  is : 
\vspace{-.2cm}
\begin{center}
$\sigma (R_p$ = 0.8775 fm) = 0.248604 mb.
\end{center}
\vspace{-.3cm}
The ratio  of the cross sections  is: 
\vspace{-.3cm}
\begin{center}
$K$ =  $\sigma(R_p$ = 0.8775 fm) / $\sigma(R_p$ = 0)  = 0.975974.    
\end{center}
\vspace{-.2cm}
Fig.\ref{fig:FigA2-1} shows the ratio  of the cross sections  $d\sigma/dt$ ($R_p$ = 0.8775 fm ) / $d\sigma/dt$ ($R_p$ = 0) generated  with the modified Dipole Form Factor. Table \ref{tabA2:TabA2-1}  presents the  results of the fits of this ratio using a polynomial $G_E = A(1 - R2Q^2 + R4Q^4- R6 Q^6 + R8Q^8)$  with fixed parameters $R6$ and $R8$.  

\vspace{1.5cm}

\begin{table}[h]
\centering
\caption{\small 
The results  of  fitting   the  ratio  of  the  differential  cross sections   $d\sigma/dt$($R_p$ = 0.8775 fm) / $d\sigma/dt$($R_p$ =0)   generated          with   the  modified  Dipole  Form  Factor.   The generated  pseudo - data  were   fitted  with  a  polynomial      function
$G_E(Q^2)  = A \cdot (1 - R2 \cdot B_2 \cdot Q^2 /C_2 + R4 \cdot B_4 \cdot  Q^4 /C_4 - \eta \cdot R_4 \cdot B_6 \cdot  Q^6 /C_6 + R8 \cdot B_8 \cdot  Q^8 /C_8) $ 
with $R8$ = 0   and various   values of fixed $R6$.   Statistics: $N_{ev}$ ($R_p$ = 0.8775 fm) = 6.96369$\cdot 10^{10}$  events.      Binning: 1000 bins.
  }
\label{tabA2:TabA2-1}
\vspace{ -0.4cm}
\begin{center}
\begin{adjustbox}{width=0.9\textwidth , center}
\begin{tabular}{ | c | c | c | c | c |c | c |} 
\hline
Fit\# & $R2$, fm$^2$ & $R2$ -- $R2$* , fm$^2$ & $R4$ , fm$^4$ & $R6$,  fm$^6$ & $R8$, fm$^8$ & $\chi^2/ndf$ \\ \hline 
\makecell{FF*  } &  0.7700* &  &     1.49* &     5.3* &  &  \\ \hline 
\makecell{Fit 1 }&  0.7681(3) &   -- 0.0019(3) &     1.33(1) &     \textbf{0} fixed &  \textbf{0}  fixed &   948/997 \\ \hline 
\makecell{Fit 2 }&  0.7700(3) &      0.0000(3) &     1.48(1) &     \textbf{5} fixed &  \textbf{0} fixed &   952/997 \\ \hline 
\makecell{Fit 3 }&  0.7720(3) &   + 0.0020(3) &     1.62(1) &    \textbf{10} fixed &  \textbf{0} fixed &   980/997 \\ \hline 
\makecell{Fit 4 }&  0.7759(3) &   + 0.0041(3) &     1.91(1) &    \textbf{20} fixed &  \textbf{0} fixed &  1110/997 \\ \hline 
\end{tabular}
\end{adjustbox}
\end{center}
\end{table}

\textit{As it follows from Table \ref{tabA2:TabA2-1}, the variation of R6 in the fitting function  from R6 = 0 to R6 = 10 fm$^6$ resulted in a systematic shift in the extracted  value of  R2 by  0.5\% (0.25\% in R$_p$). }

\end{document}